# Motion and trapping of micro- and millimeter-sized particles on the air–paramagnetic-liquid interface

Zoran Cenev[1,2], Alois Würger[3*] and Quan Zhou[1*]

[1]Department of Electrical Engineering and Automation, Aalto University, 02150 Espoo, Finland
[2]Department of Applied Physics, Aalto University, 02150 Espoo, Finland
[3]Laboratoire Ondes et Matière d'Aquitaine, Université de Bordeaux and CNRS, 33405 Talence, France

Understanding the motion of particles on an air-liquid interface can impact a wide range of scientific fields and applications. Diamagnetic particles floating on an air–paramagnetic-liquid interface are previously known to have a repulsive motion from a magnet. Here, we show a motion mechanism where the diamagnetic particles floating on the air–paramagnetic-liquid interface are attracted and eventually trapped at an off-center distance from the magnet. The behavior of magnetic particles has been also studied and the motion mechanisms are theorized in a unified framework, revealing that the motion of particles on an air–paramagnetic-liquid interface is governed not only by magnetic energy, but as an interplay of the curvature of the interface deformation created by the nonuniform magnetic field, the gravitational potential, and the magnetic energy from the particle and the liquid. The attractive motion mechanism has been applied in directed self-assembly and robotic particle guiding.

The motion of organisms on an air-liquid interface is a fundamental phenomenon in nature [1], has been profoundly studied [2], and extended to manipulate artificial objects [3]. The motion of small objects at air-liquid interfaces can be induced by surface tension gradients [4,5], magnetic fields [6–9], or perturbed interfaces from mechanical [10–12] or field stimuli [13–18]. Early work has shown that diamagnetic spherical particles on the air–paramagnetic-liquid interface can be pushed away from a magnet [19,20], attributed to negative magnetophoresis [21]. Recently, pushing non-magnetic liquid droplets and liquid marbles away from the magnet on an air–paramagnetic-liquid interface has been also demonstrated [22]. The motion of the droplets and the liquid marbles was attributed to the surface deformation of the paramagnetic liquid caused by the magnet, a phenomenon known as the inverse Moses effect [23]. These previous works have shown only pushing motion of the objects floating on the air–paramagnetic-liquid interface. However, such a prevalent notion is not complete.

In this Letter, we show that diamagnetic particles on an air–paramagnetic-liquid interface, besides being pushed, can also be pulled towards the magnet and eventually trapped at a finite distance from the centerline of the magnet. We argue that the motion mechanism is different, and it represents an interplay of the curvature of the interface deformation created by the magnet, the gravitational, and the magnetic potential of the particle and the liquid. We demonstrate that for highly concentrated paramagnetic liquids the magnetic buoyancy has a dominant role in the total energy. Additionally, when the diamagnetic particles are substituted with magnetic particles, the trapping point shifts to the peak of interface deformation induced by the magnet on the surface of the paramagnetic liquid.

We use a cylindrical rare-earth permanent magnet (25 mm × φ5 mm) to induce deformation on the air–paramagnetic-liquid interface without contacting the paramagnetic liquid. Particles floating on the interface then move towards the locations with minimum energy, which can be away from, at the base, or at the peak of the deformed interface, depending on the effective mass and the magnetic susceptibility of the particle. We have derived a model formulating the interface deformation of a paramagnetic liquid under a nonlinear magnetic field, and quantitatively analyzed the motion of the particles induced by the deformed interface experimentally and theoretically. Additionally, we have applied this pulling and trapping motion phenomenon of floating diamagnetic particles in directed self-assembly and robotic particle guiding on the surface of a magnetic medium.

In our studies, two types of paramagnetic liquids were used in all experiments: manganese dichloride-based and holmium-based aqueous solutions (see Table S1) [24]. Additionally, four types of spheroid particles were used: two types of diamagnetic particles with density similar to water, i.e., polystyrene (PS) and polyethylene (PE) particles; one type of diamagnetic low-density hollow particles, i.e., expanded polystyrene (EPS) particles; and one type of magnetic low-density particles with magnetic susceptibility higher than the paramagnetic liquids, i.e., the hollow ceramic (HC) particles (see Table S2).

Figure 1 shows the different motion modes of the particles residing on the deformed air–paramagnetic-liquid interface. Diamagnetic particles with waterlike density (PS and PE) are repelled, i.e. pushed away from the magnet [Figs. 1(a)–

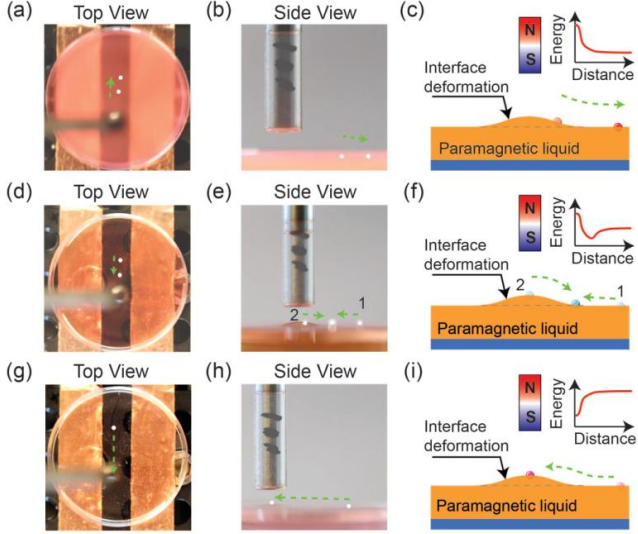

FIG. 1. Different motion modes for particles on an air–paramagnetic-liquid interface. (a) and (b), (d) and (e), (g) and (h): Experimental observations (top and side views) with (c), (f), (i) corresponding illustrations (not to scale). Green arrows denote the direction of particle motion. In (a) and (b), (d) and (e), and (g) and (h) particle visibility was enhanced with white dots. (a)–(c) A diamagnetic waterlike density PE spherical particle on the air–holmium-based paramagnetic-liquid interface is pushed. (d)–(f) A diamagnetic low-density EPS particle is pulled (1) or pushed (2) and finally trapped at the base of the interface deformation created by the magnet. (g)–(i) Magnetic low-density HC particle on the air–manganese dichloride-based paramagnetic-liquid interface is pulled and finally trapped at the peak of the meniscus. (a), (b), and (h) are superimposed images of before and after interface formation; (d) shows the initial position (1) and the trapping location of the particle; (g) shows only the initial position of the particle; (e) is an image consisting of three superimposed images, two images of starting positions (1 and 2) and one image of the final trapping location. Schematic energy profiles of the particles are shown as insets in (c), (f), and (i) where the vertical axis corresponds to the centerline of the magnet.

1(c)] on both paramagnetic liquids (Video S1). This repelling motion is in agreement with the earlier observations [19,20]. However, the diamagnetic low-density particles (EPS) are pulled or pushed by the magnet depending on their initial locations, and finally trapped at the base of the deformed interface [Figs. 1(d) and 1(f)]. One should note that this is an axisymmetric case and therefore the off-center trapping location has a ring-shaped potential well where these particles move towards. This motion has also been observed for both paramagnetic liquids (Video S2). For a magnetic low-density particle (HC), the motion is only towards the magnet with trapping location at the peak of the interface deformation [Figs. 1(g)–1(i) and Video S3]. This motion was observed for both paramagnetic liquids as well. The experimental setup for conducting the experiments is illustrated in Fig. S1(a) and detailed in the Materials and Methods section in the Supplemental Material [24].

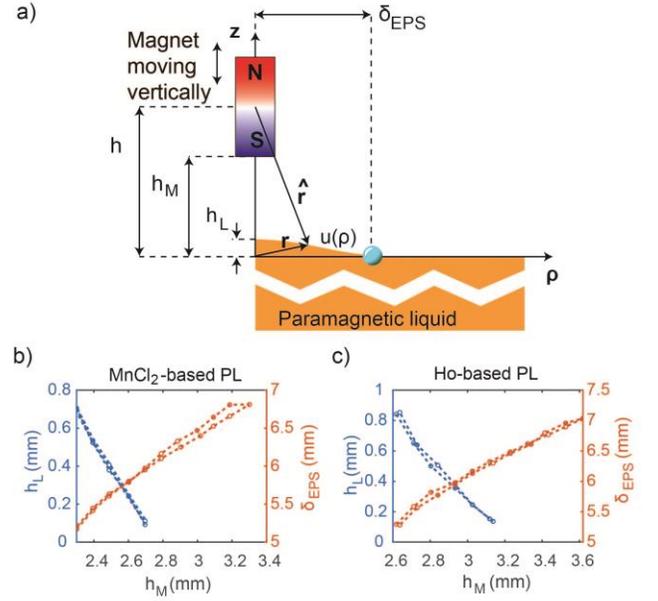

FIG. 2. Definition of parameters and effects of the air–paramagnetic-liquid interface deformation. (a) Illustration (not to scale) of system configuration showing a magnet and a deformed interface. The meniscus deformation $u(\rho)$ has its maximum value at the center, thus yielding the maximum interface deformation $h_L$. The magnet position is denoted with $h_M$. $\delta_{EPS}$ denotes the trapping location of an EPS particle with respect to the centerline of the magnet. (b) and (c) Experimental results on the $h_L$ and $\delta_{EPS}$ as a function of $h_M$ for (b) manganese dichloride-based paramagnetic liquid and (c) holmium-based paramagnetic liquid.

The relation between the vertical position of the magnet $h_M$ in respect to the maximum height of the interface deformation $h_L$ and the trapping location of a diamagnetic low-density EPS particle $\delta_{EPS}$ has been studied using the system illustrated in Fig. 2(a). Video S4 shows the experiment. The experimental results for both paramagnetic liquids are shown in Figs. 2(b) and 2(c). Both plots indicate negative correlations of the magnet position $h_M$ in respect to the maximum interface deformation $h_L$, showing the closer the magnet to the paramagnetic liquid is, the greater the interface deformation becomes. The trapping location of an EPS particle $\delta_{EPS}$ is positively correlated with the magnet-to-paramagnetic liquid distance $h_M$, i.e. a smaller the magnet to the paramagnetic liquid $h_M$ leads to a closer trapping location to the centerline of the magnet $\delta_{EPS}$, and vice versa.

The behavior of a particle on the air-liquid interface is a process of minimizing the total energy $E_{tot}$ of the system, where the particle tends to go towards regions with minimum energy. The particle interacts with both, the deformation profile of the liquid interface, and with the magnetic field. Here we briefly discuss the contributions to the total energy $E_{tot} = E_0 + E_G + E_I + E_M$, where $E_0 = \gamma \pi r_0^2 (1 - |\cos\theta|)$ is a constant representing the adsorption energy at a flat horizontal interface, with surface tension of the paramagnetic liquid $\gamma$ and a radius of the contact line $r_0 =$ $a \sin \theta$ for liquid-sphere contact angle $\theta$. With the effective mass $m_{eff}$, the gravitational acceleration $g$, and the profile of the interface deformation $u(\rho)$ as illustrated on Fig. 2(a), the potential energy reads as $E_G = m_{eff} g u(\rho)$, where $\rho$ denotes the radial distance from the centerline of the magnet. For a spherical particle of radius $a$ floating on an air–paramagnetic-liquid interface, neglecting the density of air, the effective mass is defined as $m_{eff} = \varrho_P V_P - \varrho_L V_{imm}$, where $\varrho_P$ and $V_P$ are the density and the volume of the particle, $\varrho_L$ is the paramagnetic liquid density and the $V_{imm}$ is the corresponding immersion volume $V_{imm} = \left(\frac{\pi}{3}\right) a^3 (1 + 3c_0 - c_0^3)$ with $c_0 = \cos \theta$ the degree of immersion [25–27] defined in terms of the cosine of the contact angle $\theta$ between the particle and the liquid. The experimentally measured contact angles of each particle with each paramagnetic liquid are given in Table S3 and the measurement protocol is explained in Materials and Methods section in the Supplemental Material [24].

At a non-uniform interface, the capillary energy of the particle changes due to the superposition of the deformation profile $u(\rho)$ and the meniscus induced by the effective weight $g m_{eff}$ of the particle. With the capillary length $l = \sqrt{\gamma/g\varrho_L}$ and the mean curvature of the deformed interface $H = \nabla^2 u$, the capillary energy is given by $E_I = m_{eff} g l^2 H$. The magnetic energy of the particle has two contributions, which account for its own magnetization in the field $B$ and for its magnetic buoyancy in the magnetic liquid. In the simplest case the susceptibilities are taken as constants, the magnetic energy is $E_M = B^2 \frac{1}{\mu_0}(\chi_L V_{imm} - \chi_P V_P)$, where $\mu_0$ is the permeability of free space, and $\chi_L$ and $\chi_P$ are the magnetic susceptibilities of the paramagnetic liquid and the particle, respectively. The magnetic properties ($M_v H$ curves) from which the magnetic susceptibilities $\chi_L$ and $\chi_P$ can be derived are given in Fig. S2. Finally, the total energy is an interplay among the deformation profile $u(\rho)$ and its curvature contribution $H(\rho)$ on one hand and the magnetic energy $E_M$ on the other. Therefore, we have:

$$E_{tot} = E_0 + m_{eff} g [u(\rho) + l^2 H(\rho)] \\ - (\chi_L V_{imm} - \chi_p V_p) B^2 \frac{1}{\mu_0} \quad (1)$$

The linearized curvature of the interface deformation $H(\rho)$ is given by its first and second derivatives, $H(\rho) = u''(\rho) + u'(\rho)/\rho$. The axisymmetric representation of the shape of the meniscus $u(\rho)$ varies with the distance from the centerline, as shown in Fig. 2(a) and it can be obtained by:

$$u(\rho) = \frac{1}{2\pi\gamma} \int_0^\infty dq\, q \int_0^{2\pi} d\phi\, K_0\left(\frac{|\vec{\rho} - \vec{q}|}{l}\right) \Pi(q) \quad (2)$$

where $\vec{\rho}$ is the distance along the horizontal axis, $\vec{q}$ is the sweeping vector, such that $|\vec{\rho} - \vec{q}| = \sqrt{\rho^2 + q^2 - 2\rho q \cos(\phi)}$, and $\phi$ the azimuthal angle between the two-dimensional vectors $\vec{\rho}$ and $\vec{q}$, $K_0$ is Bessel function of the second kind and $\Pi$ is the Maxwell tensor at the interface plane $z = 0$. The maximum value for $u(\rho)$ is obtained at $u(\rho = 0)$, denoted by by $h_L$ and it is directly correlated by the magnet-to-liquid distance $h_M$ as depicted in Fig. 2. A detailed theoretical derivation of the energy profiles and the interface deformation are provided in the Supplemental Material [24].

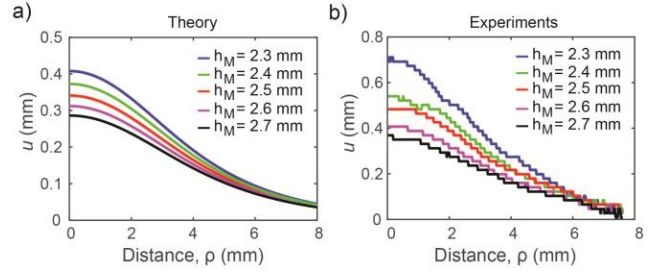

FIG. 3. Comparison of the profiles of interface deformation for manganese dichloride-based paramagnetic liquid. a) Numerical simulation and b) experimental data of $u(\rho)$ for five different vertical positions of the magnet, i.e. 2.3, 2.4, 2.5, 2.6 and 2.7 mm.

Figure 3 depicts the theoretically estimated and the experimentally obtained profiles of the interface deformation of a manganese dichloride-based paramagnetic liquid for five different magnet-to-liquid distances $h_M$, i.e. 2.3, 2.4, 2.5, 2.6 and 2.7 mm using Eq. (2). One can infer that the theoretical estimations follow the trend and the curvature of the experimentally observed ones. The comparison of the maximum height of the interface deformation $h_L$ between theoretical estimations and experimental observations can be drawn by looking at the interface deformations at distance zero, i.e. $u(0)$. For the mentioned vertical position of the magnet $h_M$ of 2.3, 2.4, 2.5, 2.6 and 2.7 mm, the theoretically estimated values for $h_L$ are 0.4, 0.37, 0.34, 0.31 and 0.28 mm and the observed values are 0.71, 0.54, 0.48, 0.40 and 0.36 mm, respectively. The maximum discrepancy of ~40% can be attributed to the linearization of interface deformation u in terms of the Young-Laplace equation, the numerical error, and the measurement error. Additional influences may also be contributed by the change of local viscosity and/or surface tension [28] in the paramagnetic liquid in the presence of magnetic field. The numerically estimated and observed profiles of interface deformation for the holmium-based paramagnetic liquid are shown in Fig. S3. The data follows the same trend and discrepancy as manganese dichloride-based paramagnetic liquid. The experimental data for both paramagnetic liquids has been extracted from Video S4.

The plots in Figs. 4(a)–4(d) show the theoretically estimated individual energy contributions of PE, PS, HC and EPS particles, respectively. The total minimum energy for the diamagnetic water-like density particles (PE and PS) is further away from the interface deformation at distances greater than 10 mm where the precise location is not indicated. This estimation agrees with the observed behavior of the particles in Video S1 and Figs. 1(a)–1(c). No trapping of particles has been theoretically estimated or observed, only pushing motion. Figure 4(c) shows the individual energy contributions for a magnetic low-density HC particle. The magnetic energy dominates both the gravitational and the capillary contributions by about three orders of magnitude. The sum of the energies has a minimum at the distance $\rho = 0$, i.e. the peak of the deformed interface. This theoretical estimation agrees with the experimental observation in Figs. 1(g)–1(i) and Video S3. Figure 4(d) shows the energetic interplay for a diamagnetic low-density EPS particle. If we ignore the local minimum from numerical error at the centerline of the magnet ($\rho = 0$), one can observe that the sum of the energy has a monotonically decreasing trend from $\rho = 0$ until $\rho \cong 7$ mm. The capillary contribution $E_I$ changes sign at $\rho \cong 4$ mm and reaches its minimum at $\rho \cong 6$ mm. The gravitational energy has a monotonically increasing trend starting from $\sim -0.6 \times 10^{-8}$ J. The sum of the three energies creates a local minimum at $\rho \cong 7$ mm. This local minimum represents the trapping location of the EPS particle. The plot in Figs. 4(d) is calculated for magnet position $h_M = 2.3$ mm.

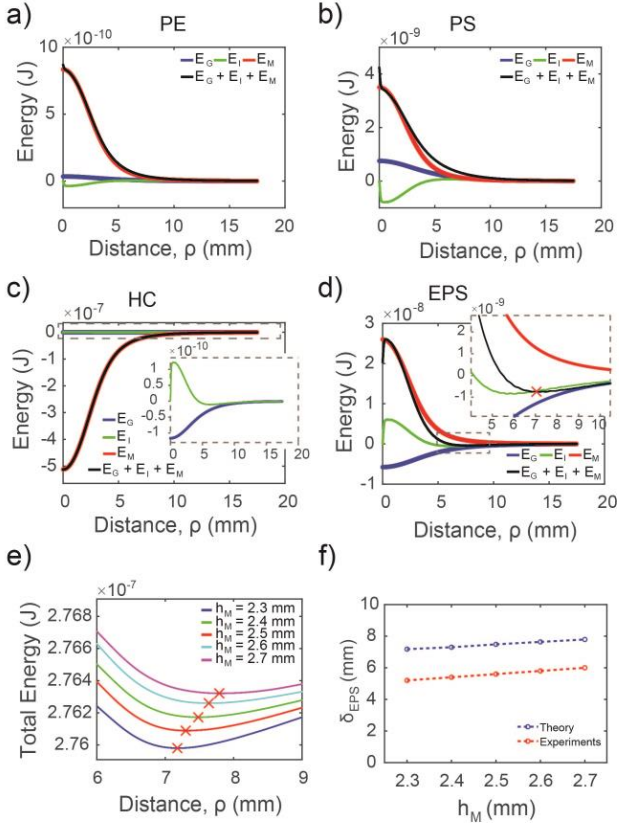

FIG. 4. Numerical estimation of energy profiles and comparisons to experimental measurements for manganese dichloride-based paramagnetic liquid. Individual energy contributions for diamagnetic waterlike density particles: (a) a PE particle and (b) a PS particle, and (c) a magnetic low-density HC particle and (d) a diamagnetic low-density EPS particle. The numerical simulations consider a magnet position of 2.3 mm in (a)–(c) and 2.6 mm in (d) (e) Numerically estimated total energy minimum, i.e., trapping location of an EPS particle at five different magnet positions $h_M$. (f) The theoretical estimation (e) compared with the experimental measurement of a trapping location of an EPS particle $\delta_{EPS}$ at five different vertical positions of the magnet (2.3, 2.4, 2.5, 2.6, and 2.7 mm). The insets in the gray dashed boxes in (c) and (d) are magnified views for the close-in plots. The red "×" signs in (d) and (e) denote energy minima.

Fig. 4(e) shows the trapping locations for five magnet positions. Fig. 4(f) compares the theoretically estimated local minima from Fig. 4(e) with the experimentally observed trapping locations from Fig. 2(b). The theoretical and experimental energy minima follow the same trend. The discrepancy in the estimated location of minimum energy is ~28%, which we attribute to the inaccuracy of the theoretical estimation of the deformed interface $u(\rho)$, the assumed constant wetting contact angle and triple contact line of EPS particle while residing on the deformed interface, and the assumed magnetic susceptibility of the EPS particles. The estimated and observed energy minima for the holmium-based paramagnetic liquid can be found in Fig. S4, where the rationale for the difference between the theoretical estimation and experimental observations is almost the same as in the case of the manganese dichloride-based paramagnetic liquid.

The plots in Figs. 4(a)–4(d) show a local minimum or maximum at the centerline of the magnet ($\rho = 0$) from the meniscus energy $E_I$. This behavior is the result of a numerical error raised in calculating the second derivative of the interface deformation $u''(\rho)$ at the boundary. Physically, the $E_I$ should have a monotonical increasing or decreasing progression at $\rho = 0$ instead of an abrupt change.

Further, the pulling motion and off-center trapping of low-density EPS particles at the base of the interface deformation on both paramagnetic liquids was observed on the replicated experimental setup as in Ref. [19] depicted in Fig. S1 (b). In this setup, a $\phi$36 mm × 4 mm disk NdFeB magnet grade N42 (supermagnete.de, Webcraft GmbH, Germany), was placed under the container filled with a paramagnetic liquid. A soft iron (a $\phi$ ~2 mm piece with low magnetic hysteresis) was installed on a motorized positioner and it was used to concentrate the magnetic field, consequently, to induce a deformation on the surface of the paramagnetic liquid. The experimental demonstrations are shown in Videos S5 and Figs. S5(b) and S5(c). We note that previous studies of negative magnetophoresis (the force resulting from the magnetic energy $F_M = -\nabla E_M$) [19,20] that describes particle motion in the absence of capillary forces and effective mass cannot explain the off-center trapped state.

To demonstrate the potential application of the pulling motion and trapping mechanism, we performed directed particle self-assembly and robotic particle guiding (Fig. 5). Four diamagnetic low-density EPS particles, initially residing on the air–paramagnetic-liquid interface, moved to the trapping ring forming a linelike bundle. When the magnet was retracted the bundle finally rearranged to an incidental T-like structure, as shown in Fig. 5(a) and Video S6. In robotic particle guiding, an EPS particle on the air–paramagnetic-liquid interface was attracted and trapped at the base of the deformation interface induced by the magnet, then the magnet executed a square trajectory with a side of 20 mm at a speed of 0.5 mm/s. The particle followed the motion of the magnet. The correlation coefficients between the X and Y coordinates of the particle and magnet were 0.90 and 0.91, respectively. This correlation indicates that the particle was indeed following the magnet as shown in Fig. 5(b) and Video S7.

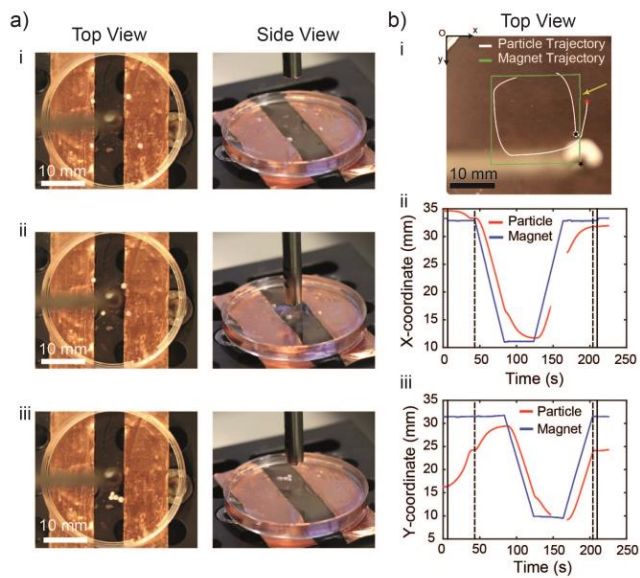

FIG. 5. Application cases of the pulling motion and trapping of the EPS particles on the air–paramagnetic-liquid interface. (a) Directed self-assembly of four EPS particles (Video S6): (i) four EPS particles residing on the air-manganese dichloride-based paramagnetic liquid interface. The magnet is perpendicular to the interface at a distance greater than 20 mm; (ii) The magnet approaches the liquid surface and forms an interface deformation which in turn pulled the particles into the trapping ring; (iii) The particles self-assembled into a line formation and after the magnet was removed the particles rearranged to a T-like structure formation. (b) Robotic particle guiding (Video S7): (i) Robotic guiding of EPS particle on the air-manganese dichloride-based paramagnetic liquid interface. The trajectories of the magnet and the particle are shown in white and green, respectively. Red stars denote start positions and black stars denote end positions. The yellow arrow shows a local error of the magnet tracking algorithm. (ii) X-coordinate of the trajectories for the particle (red) and the magnet (blue); iii) Y-coordinate of the trajectories for the particle (red) and the magnet (blue). For (ii) and (iii) the curves have missing data due to occlusion of the particle by the magnet or inadequate images for processing.

In summary, we showed that diamagnetic particles on air–paramagnetic-liquid interface, beyond being repelled, can also be attracted and trapped. The motion of diamagnetic and magnetic particles was studied to propose a unified theoretical framework for the underlying physical mechanism. The motion of a particle on the air–paramagnetic-liquid interface is treated as an energy minimization problem and is a result of an interplay among the curvature of the deformed interface created by the non-uniform magnetic field as well as the effective mass and magnetic moment of the particle. The off-center trapping location of the EPS particles varies with the vertical position of the magnet with respect to the paramagnetic liquid, the location is positively correlated with the height of the magnet. Our theoretical model correctly estimates the magnitude and the trend of the interface deformation as well as the off-center trapping location of the EPS particles. However, the model features maximum of ~40% error in the estimation the interface deformation and ~28% in the estimation of the off-center trapping location for the EPS particles. Potential applications in directed self-assembly and robotic particle guiding have also been demonstrated. The implication of this work is towards understanding the particle-liquid interactions and should inspire future research on particles interacting at the interfaces of various artificial liquids when excited by energy sources.

The authors would like also to express their gratitude to Seppo Nurmi for insightful feedback on the design and manufacturing of the mechanical parts, to Mika Latikka for experimental measurements of the magnetic properties of the paramagnetic liquids and particles as well as for lending the flat cylindrical magnet; to research scientist Dr. Michael Hummel, Dr. Meri Lundahl, Ph.D. candidates Babriela Berto and Risto Koivunen for helping in performing viscosity measurements, to SphereOne Inc. for the provision of the hollow ceramic (HC) particles-Extendospheres® 200/600, to Professor Jaakko Timonen for fruitful discussion, feedback, and contributions to the Comsol modeling, as well as to Professor Veikko Sariola for pre-reviewing this manuscript. The calculations presented in this article were performed using computer resources within the Aalto University School of Science "Science-IT" project. This research work was supported by the Academy of Finland (projects #304843 and #296250) and the Aalto University AScI/ELEC Thematic Research Programme. ZC thanks the Neles Foundation for support from a rewarded incentive stipend.

* alois.wurger@u-bordeaux.fr and quan.zhou@aalto.fi